\documentclass[osajnl,twocolumn,showpacs,superscriptaddress,10pt]{revtex4} 
\usepackage{amsmath,amssymb,graphicx,bbm}
\begin{document}

\title{Joint measurability and temporal steering}

\author{H. S. Karthik}\email{karthik@rri.res.in}\affiliation{Raman Research Institute, Bangalore 560 080, India} 
\author{J. Prabhu Tej}\email{j.prabhutej@gmail.com}
\affiliation{Department of Physics, Bangalore University, 
Bangalore-560 056, India}
\author{A. R. Usha Devi}\email{arutth@rediffmail.com} 
\affiliation{Department of Physics, Bangalore University, 
Bangalore-560 056, India}
\affiliation{Inspire Institute Inc., Alexandria, Virginia, 22303, USA.}
\author{A. K. Rajagopal} \email{attipat.rajagopal@gmail.com}
\affiliation{Inspire Institute Inc., Alexandria, Virginia, 22303, USA.}
\affiliation{Harish-Chandra Research Institute, Chhatnag Road, Jhunsi, Allahabad 211 019, India.}
\affiliation{Institute of Mathematical Sciences, C.I.T. Campus, Taramani, Chennai, 600113, India}

\begin{abstract}
Quintino {\em et. al.} (Phys. Rev. Lett.  {\bf 113}, 160402 (2014)) and  Uola {\em et. al.} (Phys. Rev. Lett.  {\bf 113}, 160403 (2014)) have recently established  an intrinsic relation between non-joint measurability and Einstein-Podolsky-Rosen steering. They showed that a set of measurements is incompatible (i.e., not jointly measurable)  if and only if it can be used for the demonstration of steering. In this paper, we prove the temporal analog of this result viz., a set of measurements are incompatible if and only if it exhibits {\em temporal steering} in a  single quantum system.   
\end{abstract}

\ocis{270.5585, 000.1600; 000.2658} 

\maketitle

\section{Introduction}

The purpose of measurements is to discern the properties of a system under investigation. In the classical scenario, all the physical observables are {\em jointly measurable} (or compatible).  In contrast, in the quantum world, non-commuting observables are  declared to be {\em incompatible} because it is not possible to  assign well-defined values to these observables jointly. The notion of {\em compatibility} of measurements is captured entirely by {\em commutativity} of the observables if one restricts only to projective valued~(PV) measurements. However, connecting compatibility of measurements with commutativity turns out to be limited in an extended framework, where the conventional idea of {\em sharp} PV  measurements of self adjoint observables gets broadened to include  generalized measurements of positive operator valued observables~\cite{Busch, BuschLahti, Lahti1, Barnett, Son, Stano, Ali,  Wolf, Yu,  Wolf2, LSW}. Active research efforts are dedicated~\cite{Busch, Barnett, Son, Stano, Ali, Wolf, Yu, Wolf2, Kar, LSW, Reeb, Kunjwal1, Kunjwal2, Brunner, Guhne, KUR}  to investigate a clear operational criteria of joint measurability within the generalized framework of POVMs and to identify the significance of {\em incompatible} measurements in revealing puzzling quantum features like Bell non-locality~\cite{Bell}, contextuality~\cite{KS} and steering~\cite{Schrodinger, Cavalcanti}. Particularly, Wolf {\em et. al.}~\cite{Wolf} have proved that the violation of the Clauser-Horne-Shimony-Holt (CHSH)~\cite{CHSH} inequality in an entangled state can be witnessed if and only if {\em incompatible} measurements of any pairs of positive operator valued measures (POVMs) with binary outcomes are employed. It has been realized that a generalized non-contextuality inequality can get violated by a quantum state in two dimensional~\cite{Kunjwalnote} Hilbert space by employing a set of three dichotomic POVMs,  which have  pairwise joint measurability  --  but no triplewise joint measurability~\cite{LSW,Kunjwal1}. In yet another recent development,  Quintino {\em et. al.}~\cite{Brunner} and Uola {\em et. al.}~\cite{Guhne} have established a general result that a set of non-jointly measurable POVMs is the one, which is useful for the task of {\em non-local steering}. 

The  concept of {\em non-local Einstein-Podolsky-Rosen (EPR) steering} was originally initiated by Schr{\" o}dinger~\cite{Schrodinger} -- as the ability to remotely prepare the state of a subsystem of an entangled state by performing local measurements on another subsystem. An experimentally testable  steering criteria was developed by Reid~\cite{Reid} for continuous variable systems (Reid proposed an uncertainty relation involving conditional variances of position and momentun violation of which demonstrates EPR non-locality in an entangled state; this is the first steering inequality though  it was not explicitly stated in Ref.~\cite{Reid}).  Wiseman {\em et. al.}~\cite{Wiseman}  formalized the task  of steering via falsification of local hidden state (LHS) models  and showed that steering constitutes a different class of non-locality that lies between entanglement and Bell non-locality.                            

The quantum steering task is usually described in terms of an example: Alice supposedly prepares a composite quantum state and sends a subsystem to Bob. She tries to convince Bob that they share an entangled state, which would allow her (with the help of local measurements on her part of the system) to remotely affect (steer) Bob's quantum state. In order to verify if Alice's claim is true,  Bob asks Alice to perform a local measurement of an observable $X_k$ on her part of the state and announce her  result $x_k$. From  local quantum state tomography (via measurements at his end), Bob determines his set of states $\{\rho_{x_k\vert k}/{\rm Tr}\,[\rho_{x_k\vert k}]\}$. If Bob's assemblage~\cite{Brunner} i.e., the set of un-normalized states $\{\rho_{x_k\vert k}\}$ (realized in each experimental run for different choices of Alice's observables $X_k$ with statistical outcomes $x_k$) does not admit a LHS decomposition of the form~\cite{lhsnote} 
\begin{equation}
\label{LHS}
\rho_{x_k\vert k}=\sum_{\lambda}\, g(\lambda)\,  p(x_k\vert k, \lambda)\, \rho_\lambda,                    
\end{equation}
(where $0\leq g(\lambda)\leq 1;\ \sum_\lambda g(\lambda)=1$ and $0\leq p(x_k\vert k, \lambda)\leq 1; \sum_{x_k}\, p(x_k\vert k, \lambda)=1$) then Bob can  convince himself that Alice is not cheating him  and they indeed share an entangled state. On the other hand, if Bob's assemblage $\{\rho_{x_k\vert k}\}$  admits a LHS structure (\ref{LHS}), he would be unable to witness violation of any of the steering inequalities~\cite{Cavalcanti} and so, concludes that Alice is fooling him. 

In order to demonstrate  steerability,  (i) Alice and Bob should necessarily share an entangled state (though the converse is not true~\cite{Wiseman}) and (ii) Alice's measurements must be comprised of incompatible POVMs~\cite{Brunner, Guhne}. Thus, in addition to entanglement being a resource for the steering task, incompatibility of measurements too plays a key role. 

Apart from the developments towards probing non-locality, in 1985 Leggett and Garg~\cite{lg} had proposed an inequality to test the concept of {\em macro-realism} in a  single quantum system in terms  of the correlations in the statistical outcomes of a dynamical observable  at different times~\cite{lgnote}. Macro-realism rests on the assumptions:  (i) physical properties of a macroscopic object exist independent of the act of observation  (postulate on the existence of  {\em reality} for all physical observables in the macroscopic world)  and (ii) measurements are {\em non-invasive} i.e., the measurements of an observable at different instants of time do not influence its subsequent dynamical evolution.   It has been experimentally demonstrated in a wide range of quantum systems~\cite{pala, mahesh, wald, souza, knee, katiyar} that {\em temporal} correlations in the outcomes of an observable  measured sequentially on a quantum system at different times do not fall under the tenet of {\em macro-realism} i.e., they violate Leggett-Garg inequality (also termed as temporal Bell inequality~\cite{Mahler, OurNote}). Very recently,  the temporal analog of  steering in a {\em single} quantum system has been proposed~\cite{KUPR, Nori}. In this paper we show that the connection between  spatial steerability and incompatible measurements of Ref.~\cite{Brunner} and \cite{Guhne} can be extended  to its temporal counterpart in a single quantum system also viz.,  non-joint measurability is necessary and sufficient for temporal steerability.   

We organize the contents of the paper as follows. We begin by giving a concise description of joint measurability of POVMs in Section~2. Section~3 is devoted to discussing temporal analog of EPR steering in a single quantum system and to establish that a set of incompatible POVMs is necessary and sufficient for temporal steering. An example to bring forth  the intrinsic connection between temporal steering and incompatibility of measurements in a single qubit system is discussed in the subsection 3.A. Concluding remarks are given in Section~4.   
       
\section{Joint measurability of POVMs} 

Mathematically, a POVM is a collection $\mathbbm{E}=\{E(x)\}$ consisting of  positive self-adjoint operators $E(x)\geq 0$ as its elements -- which sum  up to give the identity operator $\mathbbm{1}$ i.e., $\sum_x E(x)=\mathbbm{1}$. A  measurement of  $\mathbbm{E}$ in a quantum state $\rho$ results in the outcomes $x$  with probability of occurrence $p(x\vert E)={\rm Tr}[\rho\, E(x)]$. It may be noted that the POVM $\{E(x)\}$ encompasses the conventional PV measurements as a special case (when the elements $E(x)$ of the POVM  constitute a complete and orthogonal set $\{\Pi(x)\}$ of projectors). 

In the restricted measurement scenario, where only PV measurements are considered, it is well established that commuting physical observables are jointly measurable. But in the extended framework of generalized measurements, the notion of joint measurability is non-trivial and has received increasing attention~\cite{Busch, BuschLahti, Lahti1, Barnett, Son,  Stano, Ali, Wolf, Yu, Wolf2, LSW}. A more refined notion of compatible (jointly measurable) POVMs is defined as follows. A set of POVMs $\mathbbm{E}_k=\{E_k(x_k)\}$ is said to be compatible  if the probabilities $p(x_k\vert k)={\rm Tr}\,[\rho\, E_k(x_k)]$  of their outcomes  in any arbitrary quantum state $\rho$  can be discerned by measuring a {\em global} POVM $\mathbbm{G}=\{G(\lambda); G(\lambda)\geq 0,\,\sum_\lambda\, G(\lambda)=\mathbbm{1}\}$ -- where the measurement  outcomes  $\lambda=\{x_1,x_2,\ldots\}$ of $G(\lambda)$ occur with probabilities $g(\lambda)={\rm Tr}\, [\rho\, G(\lambda)]$ -- by classical post processing of the data~\cite{Ali, Guhne}:   
\begin{equation}
\label{pxilambda}
p(x_k\vert k)=\sum_{\lambda} g(\lambda)\,  p(x_k\vert k, \lambda) 
\end{equation} 
where $0\leq p(x_k\vert k, \lambda)\leq 1$ in (\ref{pxilambda}) are some arbitrary positive numbers satisfying $\sum_{x_k}\, p(x_k\vert k, \lambda)=1$.   

More precisely, associated with a set of jointly measurable POVMs $\{E_k(x_k)\}$ there exists a  grand POVM $\mathbbm{G}=\{G(\lambda)\}$ such that  
\begin{equation}
\label{masterG}
E_k(x_k)= \sum_{\lambda} p(x_k\vert k, \lambda)\, G(\lambda)\ \ \  \forall \ \  k. 
\end{equation}
In other words, it suffices to measure the grand POVM $\mathbbm{G}$ to discern the measurement results of compatible POVMs $\mathbbm{E}_k$. 

An important aspect to be  highlighted here is that the generalized POVMs are jointly  measurable even if they do not commute with each other. 
      
Consider a triad of qubit observables $X=\vert 0\rangle\langle 1\vert+\vert 1\rangle\langle 0\vert$, 
$Y=-i\, \vert 0\rangle\langle 1\vert+\, i\, \vert 1\rangle\langle 0\vert$ and $Z=\vert 0\rangle\langle 0\vert-\vert 1\rangle\langle 1\vert$ measured by employing the POVMs $\mathbbm{E}_{X},\, \mathbbm{E}_{Y}$ and $\mathbbm{E}_{Z}$ defined in terms of their elements      
\begin{eqnarray}
\label{unsharpxyz}
E_X(x)&=& \frac{1}{2}\left(\mathbbm{1}+ \eta\, x\, X \right),\nonumber \\
E_Y(y)&=& \frac{1}{2}\left(\mathbbm{1}+ \eta\, y\,Y \right),\nonumber \\
E_Z(z)&=& \frac{1}{2}\left(\mathbbm{1}+ \eta\, z\, Z \right).
\end{eqnarray} 
The measurements result in binary outcomes  $x,\, y,\, z=\pm 1$ and they correspond to  {\em fuzzy} measurements of the observables $X,\, Y,\, Z$,  characterized by the {\em unsharpness} parameter $0\leq \eta\leq 1$. It has been identified that the  qubit POVMs $\{E_{X}(x)\}$, $\{E_{Y}(y)\}$  and $\{E_{Z}(z)\}$ are {\em pairwise} jointly measurable if and only if $\eta \leq 1/\sqrt{2}$ and the condition $\eta\leq 1/\sqrt{3}$ is necessary and sufficient for their triplewise joint measurability~\cite{Stano, LSW, Kunjwal1}. It may also be noted that  when $\eta=1$, the POVMs $\{E_{X}(x)\}, \{E_{Y}(y)\}, \{E_{Z}(z)\}$ reduce to their corresponding {\em sharp} PV versions $\{\Pi_{X}(x)\}, \{\Pi_{Y}(y)\}, \{\Pi_{Z}(z)\}$.  

\section{Temporal steering and incompatible measurements}

We consider a system prepared in a quantum state $\rho=\rho(0)$, which evolves under the Hamiltonian evolution $U(t)=e^{-i\,H\, t/\hbar}$, dynamically transforming the state (in the Schrodinger picture) as $\rho\rightarrow \rho(t)=U(t)\, \rho\, U^\dag(t)$ at time $t$. The physical observables undergo dynamical evolution (in the Heisenberg picture)  as  $X(0)\rightarrow X(t)=U^\dag(t)\, X(0)\, U(t)$. The observable  $X$ at different time instants $t_k$ (which we denote by $X_k$) do not commute in general.  Hence $\{X_k\}$ are not {\em jointly measurable} within the restricted framework of PV measurements. Contrast this situation with the classical scenario, where measurement of an observable at a given instant of time does not disturb its subsequent evolution. In other words, one can measure an observable at different instants of time jointly in the classical scenario. In the quantum case,  measurements  of non-commuting observables, in general, form an incompatible set of measurements.  

To illustrate the temporal analog of steering, we consider a game involving two players Alice and Bob. Alice prepares a state $\rho$ (which is not disclosed to Bob).  Bob asks Alice to measure the observable $X$ at different instants of time $t_k$ using incompatible POVMs. Alice claims that she has measured $X_k$ and obtained an outcome $x_k$ with probability $p(x_k\vert k)$. She gives the post measured states  $\frac{\rho_{x_k\vert k}}{{\rm Tr}[\rho_{x_k\vert k}]}$ to Bob. 
Bob's task is to verify if Alice has given him a genuine set of post measured assemblage $\{\rho_{x_k\vert k}\}$,  where the unnormalized states
\begin{equation} 
\rho_{x_k\vert k}=\sqrt{E_{k}(x_k)}\, \rho\,\, \sqrt{E_{k}(x_k)}
\end{equation} 
have resulted via  measurements of incompatible POVMs $\{E_{k}(x_k)\}$ of the observables $X_k$  or if Alice is cheating him by merely stating that she has performed the measurements. 
 
Bob can only trust his measurements on the states handed over to him by Alice, with a pre-label $\{x_k, p(x_k\vert k)\}$. In order to accomplish the protocol, Bob may choose to measure the observables $X_l$ at a later time,  $l\geq k$ on the assemblage $\rho_{x_k\vert k}$ and record the conditional probabilities  ${\cal P}(x_l\vert x_k)$ of his outcomes $x_l$ (given that Alice had obtained an outcome $x_k$ in her measurement of the observable $X_k$); he then explores if the {\em temporal correlations} of the observables $X_k$, $X_l$ violate any steering inequality~\cite{Note2}. If the temporal steering inequality is violated, then Bob concludes that Alice has indeed performed incompatible measurements of the observables $X_k$.  We refer to this scenario as {\em temporal steering}~\cite{Note3}.   

More generally, Bob could determine the assemblage $\{\rho_{x_k\vert k}\}$ given to him through quantum state tomography; if the assemblage $\rho_{x_k\vert k}$ is of the {\em hidden state}~(HS) form, $\rho_{x_k\vert k}=\sum_{\lambda}\, g(\lambda)\,  p(x_k\vert k, \lambda)\, \rho_\lambda$ (which is exactly identical to the LHS form (\ref{LHS}))  where $0\leq p(x_k\vert k,\lambda)\leq 1,\ \sum_{x_k}\, p(x_k\vert k,\lambda)=1$,  then  Bob convinces himself that the assemblage $\{\rho_{x_k\vert k}\}$ is not {\em temporally steered}.  Because the actual scenario may be the following.  Alice could have drawn some random states $\rho_\lambda$ with probability $g(\lambda)$ (from a statistical mixture $\rho=\sum_\lambda\, g(\lambda)\, \rho_\lambda$)  --  but  announce that an outcome $x_k$ has occurred in the measurement of the observable $X$ at time $t_k$, with a probability of occurrence $p(x_k\vert k,\lambda)$ (Alice could  have theoretically calculated the  probabilities $p(x_k\vert k,\lambda)$ for the hypothetical outcomes $x_k$ of measurement).  If Alice has indeed performed incompatible measurements, as she claims,  Bob's assemblage $\{\rho_{x_k\vert k}\}$ deviates from the HS form. Bob can then convince himself that Alice has indeed given  him  a set of states which reveals {\em temporal steering}, and it has resulted from the measurements of the observable $X$ at different instants of time using  incompatible POVMs.   

We now proceed to show that measurements of $\{X_k\}$ using a {\em compatible} set $\{\mathbbm{E}_k\}$ of POVMs  do not lead to temporal steering. 

Let us suppose that Alice performs measurement of a global POVM $\mathbbm{G}=\{G(\lambda)\}$. After her measurement  the post measured states are given by 
\begin{equation}
\rho_\lambda=\sqrt{G(\lambda)}\, \rho\, \sqrt{G(\lambda)}/g(\lambda),
\end{equation} 
where $g(\lambda)={\rm Tr}\,[\rho\, G(\lambda)]$ is the probability of outcome $\lambda$.  Alice would then classically post process the measurement data of the global POVM  $\mathbbm{G}=\{G(\lambda)\}$  to obtain the probabilities of outcomes $p(x_k\vert k)$ of measurement of any compatible POVMs $\mathbbm{E}_k$ to have resulted in  an  outcome $x_k$ as, 
\begin{eqnarray} 
\label{pxk}
p(x_k\vert k) &=&{\rm Tr}\,[\rho\, E_k(x_k)]\nonumber \\ 
&=&\sum_{\lambda}\, p(x_k\vert k,\lambda)\, {\rm Tr}[\rho\, G(\lambda)] \nonumber \\
&=& \sum_{\lambda}\,   p(x_k\vert k,\lambda)\, g(\lambda).
\end{eqnarray} 
More specifically, Alice could discern the results of measurements of {\em compatible} POVMs $\mathbbm{E}_k=\{E_k(x_k)\}$ via measurement of a global POVM $\mathbbm{G}=\{G(\lambda)\}$ and then using the decomposition $E_k(x_k)=\sum_\lambda\, p(x_k\vert k, \lambda)\, G(\lambda)$.  

After Alice announces her  measurement results $\{x_k,\, p(x_k\vert k)\}$ of $E_k(x_k)$ and hands over the post measured set of states, Bob detects  that  his assemblage $\{\rho(x_k\vert k)\}$ is of the HS form  $\rho(x_k\vert k)=\sum_\lambda\, g(\lambda)\, p(x_k\vert k, \lambda)\, \rho_\lambda$.   
Thus,  Bob concludes that there is no temporal steering.   
 
Conversely, we prove that non-jointly measurable (incompatible) POVMs  are sufficient to demonstrate  temporal steering. 
To bring this out, we consider a completely random state $\rho=\mathbbm{1}/d$ and a set of POVMs $\{\mathbbm{E}_k\}$ for the  measurements of the observables $\{X_k\}$. The  post measured assemblage  $\{\rho_{x_k\vert k}\}$ is characterized by its elements, 
\begin{eqnarray} 
\rho_{x_k\vert k}&=&\sqrt{E_k(x_k)}\, \rho\, \sqrt{E_k(x_k)} \nonumber \\
&=& \frac{1}{d}\, E_k(x_k). 
\end{eqnarray}         
One can thus express the elements $E_k(x_k)$ of the POVM in terms of the assemblage  $\{\rho_{x_k\vert k}\}$  as~\cite{noteproof},
\begin{equation}
 E_k(x_k)=d\, \rho_{x_k\vert k}
\end{equation}
If there is no temporal steering, then the assemblage $\{\rho_{x_k\vert k}\}$ is described by a HS form (\ref{LHS}) and hence one obtains 
\begin{eqnarray}
\label{comp}
E_k(x_k)&=&d\, \sum_{\lambda}\, g(\lambda)\, p(x_k\vert k,\lambda) \rho_\lambda \nonumber \\
&=& \sum_{\lambda}\, p(x_k\vert k,\lambda) G(\lambda)
\end{eqnarray} 
where  $G(\lambda)=d\,\, g(\lambda)\, \rho_\lambda$.  It is easy to see that  (\ref{comp}) is essentially the joint measurability condition (see (\ref{masterG})) for the definition of compatible POVMs $\{E_k(x)\}$. We thus obtain the result: {\em A set of POVMs is said to be non-jointly measurable if and only if it is useful for demonstrating temporal steering}. 

Our result highlights that one does not require a steerable entangled state to determine if a given set of measurements is compatible or not;  it suffices to detect temporal non-steerability  in a single quantum system itself to accomplish this task. 

\subsection{Joint measurability and temporal steering in a single qubit system}

For the purpose of illustrating the intrinsic connection between temporal steering and non-joint measurability of the POVMs in a single qubit system, suppose that Alice prepares a single qubit system in a maximally disordered state $\rho=\frac{\mathbbm{1}}{2}$. Bob asks Alice to subject the system to a Hamiltonian evolution $U(t)=e^{-i\, H\, t/\hbar}$ where the Hamiltonian $H=\hbar\, \omega\, Z= \hbar\, \omega\,  (\vert 0\rangle\langle 0\vert-\vert 1\rangle\langle 1\vert)$ and measure the observables $X_k =U^\dag(t_k)\, X\, U(t_k)= X\, \cos(\omega\, t_k) + Y\, \sin(\omega\, t_k)$;\ $X=\vert 0\rangle\langle 1\vert+\vert 1\rangle\langle 0\vert$,\ \ $Y=-i\, (\vert 0\rangle\langle 1\vert-\vert 1\rangle\langle 0\vert)$ at two different time intervals (i) $t_1=0$ and (ii) $t_2=\pi/(2\,\omega)$  using incompatible POVMs. Alice employs binary outcome POVMs $\{E_X(x)=\frac{1}{2}\left(\mathbbm{1}+ \eta\, x\, X \right),\ x=\pm 1\}$ and $\{E_Y(y)= \frac{1}{2}\left(\mathbbm{1}+ \eta\, y\, Y \right),\ y=\pm 1\}$ to measure the observables $X_1=X$, $X_2=Y$ respectively. After her measurements on several identically prepared  copies of the initial state, Alice hands over four different assemblages $\{\rho_{x=\pm 1\vert 1}\}$, $\{\rho_{y=\pm 1\vert 2}\}$ -- labelled by the outcomes of measurements $x=\pm 1$, $y=\pm 1$  and  the corresponding  probablilities of occurrence  $p(x=\pm 1\vert 1),\ p(y=\pm 1\vert 2)$ to Bob. Bob then chooses to perform PV measurement $\{\Pi_X(x')= \frac{1}{2}\left(\mathbbm{1}+ x'\, X \right); x'=\pm 1\}$ at time   $t_{3}=2\,\pi/\omega$ on the assemblage $\{\rho_{x=\pm 1\vert 1}\}$ (note that at $t_{3}=2\,\pi/\omega$,  the  observable  $X_3=X$);  he obtains the conditional probabilities for his measurement outcomes $x'=\pm 1$:  
\begin{eqnarray} 
\label{cx}
{\cal P}(x'\vert x)&=&{\rm Tr}[\rho_{x\vert 1}\, \Pi_X(x')]/p(x) \nonumber \\ 
&=&  {\rm Tr}[E_X(x)\, \Pi_X(x')] \nonumber \\
&=& \frac{1}{2}\, (1+\eta\, x\, x').  
\end{eqnarray}
Further, Bob carries out PV measurements $\{\Pi_Y(y')= \frac{1}{2}\left(\mathbbm{1}+ y'\, Y \right); y'=\pm 1\}$ at time $t_{4}=5\,\pi/(2\,\omega)$ on the assemblage $\{\rho_{y=\pm 1\vert 1}\}$ (at $t_{4}= 5\,\pi/(2\, \omega)$,  the  observable  $X_4=Y$) and registers the conditional probabilities for his measurement outcomes $y'=\pm 1$:
\begin{eqnarray}
\label{cy} 
{\cal P}(y'\vert y)&=&{\rm Tr}[\rho_{y\vert 2}\, \Pi_Y(y')]/p(y), \nonumber \\ 
&=&  {\rm Tr}[E_Y(y)\, \Pi_Y(y')] \nonumber \\
&=& \frac{1}{2}\, (1+\eta\, y\, y').  
\end{eqnarray} 
(In the second lines of (\ref{cx}), (\ref{cy}), we have substituted the probabilities of Alice's outcomes $p(x)={\rm Tr}[\rho\, E_X(x)]=1/2$, $p(y)={\rm Tr}[\rho\, E_Y(y)]=1/2$ in  the qubit state $\rho=\frac{\mathbbm{1}}{2}$.) 

As the expectation values of the qubit observables $(X+Y)/\sqrt{2}$ in any arbitrary qubit state is constrained to be less than 1 (the maximum eigenvalue of the observable), evidently the conditional expectation value of the observable $(X+Y)/\sqrt{2}$  (evaluated from Bob's  measurement outcomes $x'$, $y'$  of $X$, $Y$   --  which are conditioned by Alice's POVM outcomes  $x$, $y$ of the same observables) too is restricted i.e.,  
\begin{eqnarray}
\left\langle \frac{(X + Y)}{\sqrt{2}}\right\rangle_{x, y}= \frac{1}{\sqrt{2}}\, \left(\langle X\rangle_{x}+\langle Y\rangle_{y}\right)\hskip 1in \nonumber \\
=\frac{1}{\sqrt{2}}\, \left[\sum_{x'}\, {\cal P}(x'\vert x)\, x' + \sum_{y'}\, {\cal P}(y'\vert y)\, y'\right]\leq 1.\nonumber \\
\end{eqnarray}

If the assemblages $\{\rho_{x=\pm 1\vert 1}\}$, $\{\rho_{y=\pm 1\vert 1}\}$, obtained  after Alice performs her measurements, constitute a HS structure i.e., $\rho_{x=\pm 1\vert 1}=\sum_{\lambda}\, g(\lambda)\, p(x\vert 1; \lambda) \,\rho_\lambda$ and $\rho_{y=\pm 1\vert 1}=\sum_{\lambda}\, g(\lambda)\, p(y\vert 1; \lambda) \,\rho_\lambda$, then Bob's measurements lead to a linear temporal steering inequality (obtained following the arguments outlined by Cavalcanti {\em et. al.}~\cite{Cavalcanti} for the derivation of linear EPR steering criteria for two spatially separated qubits):    
\begin{equation}
\label{linT}
\left\vert \,\,\sum_{x=\pm 1}\, p(x)\, x\, \langle X\rangle_x + \sum_{y=\pm 1}\, p(y)\, y\, \langle Y\rangle_y \, \right\vert\leq \sqrt{2} 
\end{equation}
Substituting the conditional probabilities (\ref{cx}) and (\ref{cy}) to evaluate the expectation values $\langle X\rangle_x$, $\langle Y\rangle_y$ and simplifying,   the linear temporal inequality (\ref{linT}) results in the constraint 
\begin{equation}
\eta \leq \frac{1}{\sqrt{2}}  
\end{equation}  
on the unsharpness parameter --    
which is exactly the condition for joint measurability~\cite{Stano, LSW, NOTE} of the qubit observables $X$, $Y$ using the POVMs $\{E_X(x)=\frac{1}{2}(\mathbbm{1}+\eta\, x\, X);\, x=\pm 1\}$ and $\{E_Y(y)=\frac{1}{2}(\mathbbm{1}+\eta\, y\, Y);\, y=\pm 1\}$. The temporal steering inequality (\ref{linT}) is violated for $\frac{1}{\sqrt{2}}<\eta\leq 1$ i.e., when Alice's POVMs are incompatible.

\section{Conclusions} 

We have illustrated temporal steering phenomena in a single quantum system by developing the notion of a HS structure -- which is analogous to the LHS model for spatially separated systems.  Falsification of HS model implies temporal steerability. Extending the arguments of recent papers \cite{Brunner, Guhne}  we have established a relation between incompatibility of quantum measurements and temporal steering phenomena. Our results highlight that a set of measurements are incompatible if and only if they can be used to demonstrate temporal steering in any quantum state. The connection between measurement incompatibility and temporal steering opens up new avenues for exploring temporal steering inequalities to infer about (non) joint measurability. Further, following similar lines of investigations on non-local steering vs Bell non-locality~\cite{Wiseman} of spatially separated states, it would be of  interest to investigate if Leggett-Garg inequalities and temporal steering inequalities carry identical inferences about measurement invasiveness~\cite{note_invasive} or if they bring forth its different manifestations. We leave open these aspects for future investigations.

\noindent {\bf Acknowledgements:} One of us (JP) acknowledges support from UGC-BSR, Government of India.


\begin{thebibliography}{99}
\bibitem{Kunjwalnote} 
Conventionally, the notion of quantum contextuality~\cite{KS}) is demonstrated by using projective measurements in Hilbert spaces of dimension $d\geq 3$.
\bibitem{lhsnote} 
It could be that  Alice randomly draws  states $\rho_{\lambda}$ with probability $g(\lambda)$ from a chosen ensemble  and sends it to Bob. And when Bob asks her to perform measurement of the observable $X_k$,  she could communicate a fake outcome $x_k$ to have occurred with the probability  $p(x_k\vert k)=\sum_\lambda\, g(\lambda)\, p(x_k\vert k,\lambda)$ (based on  her knowledge of a pre-existing LHS model). In this scenario, Bob could  detect that steering inequalities~\cite{Cavalcanti} are not violated  -- which in turn reveals that  LHS model (\ref{LHS}) indeed holds.
\bibitem{noteproof} Note that $\rho_{x_k\vert k}=\frac{1}{d}\, E_{k}(x_k)>0$ and the set of operators $\{d\, \rho_{x_k\vert k}\}$ obey
$\sum_{k} d\, \rho_{x_k\vert k}= \sum_k E_{k}(x_k)= \mathbbm{1}$.
\bibitem{NOTE}
Necessary and sufficient conditions for the joint measurability of a generic set of $N$ dichotomic qubit POVMs $\{E_k(x_k),\ x_k=\pm 1;  k=1,2,\ldots, N\}$  have been derived  in Ref.~\cite{LSW, Kunjwal1}, by identifying the range of the unsharpness parameter $\eta$ for which the POVMs are compatible/incompatible.
\end{thebibliography}
\end{document}